\documentclass[journal,letterpaper,onecolumn,draftclsnofoot,12pt]{IEEEtran}
\usepackage{cite}

\ifCLASSINFOpdf
\else
   \usepackage[dvips]{graphicx}
\fi
\usepackage[cmex10]{amsmath}
\usepackage{array}

\usepackage[tight,footnotesize]{subfigure}

\usepackage{stfloats}
\usepackage{url}

\usepackage{psfrag}

\usepackage{dsfont}

\newcommand{\argmin}{\operatornamewithlimits{argmin}}
\newcommand{\argmax}{\operatornamewithlimits{argmax}}

\usepackage[normalem]{ulem}

\usepackage{algpseudocode}
\usepackage{algorithm}

\usepackage{amsfonts}

\newcommand{\E}[1]{{{\rm E} \left\{ #1 \right\}}}
\newcommand{\tr}[1]{{{\rm tr} \left( #1 \right)}}

\renewcommand{\log}[1]{{\rm log}_{2}\left({#1}\right)}
\renewcommand{\ln}[1]{{\rm ln}\left({#1}\right)}
\newtheorem{mydef1}{Proposition}

\begin{document}
%
\title{Maximizing the Sum Rate in Cellular Networks Using Multi-Convex Optimization}
%
%
%

\author{Hussein~Al-Shatri,~\IEEEmembership{Member,~IEEE,}
        Xiang~Li,~\IEEEmembership{Student Member,~IEEE,}
        Rakash~SivaSiva~Genesan,~\IEEEmembership{Student~Member,~IEEE,}
        Anja~Klein,~\IEEEmembership{Member,~IEEE,}
        and~Tobias~Weber,~\IEEEmembership{Senior~Member,~IEEE}\thanks{This work has been submitted to the IEEE for possible publication. Copyright may be transferred without notice, after which this version may no longer be accessible.}}

\maketitle

\begin{abstract}
In this paper, we propose a novel algorithm to maximize the sum rate in interference-limited scenarios where each user decodes its own message with the presence of unknown interferences and noise considering the signal-to-interference-plus-noise-ratio. It is known that the problem of adapting the transmit and receive filters of the users to maximize the sum rate with a sum transmit power constraint is non-convex. Our novel approach is to formulate the sum rate maximization problem as an equivalent multi-convex optimization problem by adding two sets of auxiliary variables. An iterative algorithm which alternatingly adjusts the system variables and the auxiliary variables is proposed to solve the multi-convex optimization problem. The proposed algorithm is applied to a downlink cellular scenario consisting of several cells each of which contains a base station serving several mobile stations. We examine the two cases, with or without several half-duplex amplify-and-forward relays assisting the transmission. A 
sum power constraint at the base stations and a sum power constraint at the relays are assumed. Finally, we show that the proposed multi-convex formulation of the sum rate maximization problem is applicable to many other wireless systems in which the estimated data symbols are multi-affine functions of the system variables.
\end{abstract}
\begin{IEEEkeywords}
sum rate maximization, interference, multi-convex function, amplify-and-forward relay.
\end{IEEEkeywords}

%
\IEEEpeerreviewmaketitle

\section{Introduction}
\label{sec:1}
\IEEEPARstart{S}{everal} sophisticated solutions have been studied for future cellular systems aiming at improving both the uplink and downlink data rates. For instance, introducing multiple antennas at both base stations (BSs) and mobile stations (MSs) greatly increases the achievable rates \cite{F96,FG98,T99}. Furthermore, employing relays in these systems extends the coverage and enhances the performance \cite{RW07,BKWUK09}. However, interference is still the main performance limiting factor in cellular systems. The transmission rate, especially when the MSs are located at the cell edges, is greatly influenced by the inter-cell interferences. For instance for a cell edge MS, the received interference signal in the downlink can be severe and even of a comparable strength as the useful signal, which degrades the achieved rate significantly. To enhance the performance in cellular systems, smart spatial signal processing techniques at the BSs and the MSs, and also at the relays if they are employed in the 
system, need to be found. Apart from joint processing techniques which require data exchange among the cooperating parties, we focus on distributed signal processing techniques which require only the exchange of channel state information.

Before discussing the sum rate, we first briefly review two interference reduction techniques which have been studied extensively, i.e., interference alignment (IA) and sum mean square error (MSE) minimization. IA is achieved by aligning all the interferences in a smaller subspace of the received signal space while keeping the useful signal subspace interference free \cite{JS08,CJ08}. IA has received great attention in the last few years \cite{GCJ11,AW11,PH09,TGR09,GWK11}. Basically, the IA problem has the nice property that it is a multi-affine problem. Therefore, it can be tackled by alternatively solving several linear subproblems \cite{PH09,GCJ11}. For instance, the IA problem is a tri-affine problem if relays are employed. Firstly, the filters of the BSs are optimized with fixed relay processing matrices and fixed filters at the MSs. Secondly, the relay processing matrices are optimized with fixed filters at the BSs and MSs. At the third step, the filters of the MSs are optimized with fixed filters at 
the BSs and fixed relay processing matrices. However, since IA ignores the received noise, it performs poorly at low and moderate signal to noise ratios (SNRs) \cite{EAPH13}. On the other hand, optimizing the spatial filters at the BSs and the MSs, as well as the processing matrices at the relays if they are employed, for minimizing the sum MSE always achieves a compromise between interference reduction and noise reduction. In general, the sum MSE is not a convex function. However, it is a convex function of either the filters at the BS, the filters at the MSs or the relay processing matrices alone. This multi-convex structure of the sum MSE function also allows alternating minimization algorithms to achieve a local minimum \cite{SSBHU09,SLTW10,MXFWNP10,GAWK13}. Nevertheless, minimizing the sum MSE does not necessarily imply achieving the maximum sum rate. Furthermore, it is worth to mention here that minimizing the sum bit error rate (BER) is an alternative objective to minimizing the sum MSE \cite{CAH05}. 
However, it is complicated to optimize the sum BER in multi-user multi-antenna scenarios as the BER has no closed form solution.

Besides the techniques mentioned above, directly maximizing the sum rate is a promising goal for efficiently utilizing the limited available system resources \cite{S95}. If the interference is treated as noise and some power constraints are considered, the sum rate maximization problem is a non-convex optimization problem \cite{LZ08,HL09}. This non-convexity of the sum rate maximization problem holds even if we optimize over either the filters at the BSs only, the filters at the MSs only or the relay processing matrices only. Therefore, iterative alternating optimization algorithms cannot be directly implemented here.

In the last decade, a lot of progress has been made in finding efficient sum rate maximization algorithms. Algorithms from global optimization theory are proposed for finding the global maximum of the sum rate \cite{AW12,QZH09,JL10}. Nevertheless, these algorithms suffer from high computational complexity which limits their practicality to small scenarios only. Unlike the computationally expensive global optimization algorithms, relatively low complexity suboptimum algorithms have also received great attention. Basically, the special structure of the sum rate function can be exploited to achieve a near optimum sum rate. In \cite{CTJL07}, an interference broadcast channel is considered. Instead of maximizing the sum rate, the authors maximizes the product of the SNRs at the MSs. Rather than optimizing the filters at the BS and the MSs all together, it is shown that the problem can be simplified to three subproblems, which are not necessarily convex. Each subproblem aims at optimizing either the transmit 
powers, the BS filters or the MS filters. Geometric programming is employed for approximating the solution of the non-convex subproblems. In \cite{ZXL09}, some auxiliary variables are used to simplify the sum rate maximization problem in a broadcast channel. The authors introduce new variables to the problem such that the multiple constraints can be equivalently written as a single constraint. The sum rate function can be written as a difference of two concave functions \cite{VSS10,AW12}. Accordingly, the authors of \cite{VSS10} linearly relax the second term and solve the resulting problem iteratively.

Some authors also exploit the minimized MSE to maximize the sum rate. From the information theory perspective, Guo \emph{et.} \emph{al.} have found that there is a linear relationship between the derivative of the mutual information and the minimum MSE for Gaussian channels \cite{GSV05}. Moreover, it is shown in \cite{PV06} that this relationship holds for any wireless system with linear filters. Considering a broadcast channel scenario, the relationship between the derivative of the mutual information and the minimum MSE can be exploited by designing the receive filters such that the MSE at the receivers is minimized. In this case, the MSE will be a function of the transmit filters \cite{CACC08}. Accordingly, the sum rate maximization problem for optimizing the transmit filters can be formulated as a minimization of the sum of log-MSEs. An approximate solution of this new formulation is found using geometric programming \cite{SSB08}. Designing the receive filters to minimize the MSE and optimizing the 
remaining variables to maximize the sum rate is also considered in \cite{KTJ12}. It is shown that by relaxing the sum rate maximization problem and adding some auxiliary variables, a successive convex approximation approach can be applied \cite{KTJ12}. Because of the problem relaxations, this approach does not converge to neither a local maximum nor the global maximum of the original problem. For a broadcast channel scenario, the receive filters are designed aiming at minimizing the MSE and by adding some auxiliary variables, the sum rate maximization problem is reformulated as a biconvex optimization problem of the transmit filter and the added auxiliary variables \cite{CACC08}. This work is extended to many different scenarios such as MIMO interference channels \cite{NSGS10}, interfering broadcast channels \cite{SRLH11}, and relay interference channels \cite{THJ11}. The main drawback of this approach is that the receive filters are not optimized to maximize the sum rate.

In the present paper, we aim at formulating the sum rate maximization problem as a multi-convex problem so that it can be efficiently solved by low complexity iterative algorithms. We specifically consider the downlink transmission in a cellular scenario with BSs serving multiple MSs, although the same approach can be applied to the uplink transmission as well. The transmission from the BSs to the MSs takes place either through several non-regenerative relays or directly without relays. First, we focus on describing our approach for a two-hop transmission scheme where relays are employed. Then, we show that the approach can also be applied to other wireless systems by taking the single-hop transmission scheme without relays as an example. The key idea of our approach is to replace the signal to interference plus noise ratio (SINR) at a MS by a new term whose maximal value is found to be 1+SINR. Using this new term, we formulate a multi-concave objective function. We will show that this objective function has 
the same maxima as the sum rate function and, therefore, maximizing this objective function is equivalent to maximizing the sum rate function.

The rest of this paper is organized as follows. In the next section, a two-hop transmission scenario and a single-hop transmission scenario are described. In Section \ref{sec:5}, the two-hop transmission is first investigated and based on it, the multi-convex formulation of the sum rate is derived. An iterative sum rate maximization algorithm is proposed in Section \ref{sec:5.5}. To show that our idea is quite general and fits in many scenarios, we derive the multi-convex problem formulation for the single-hop transmission in Section \ref{sec:6}. A few additional aspects are discussed in Section \ref{sec:disc} and the performance of the proposed algorithm is shown in Section \ref{sec:7}. In Section \ref{sec:8}, the conclusions are drawn.

\section{System model}
\label{sec:3}
\subsection{Two-Hop Interference Broadcast Scenario} \label{sec:2.1}
In this paper, we will consider two related scenarios, i.e., a two-hop interference broadcast scenario and a single-hop interference broadcast scenario. The former will be described here, and the latter will be described in Section \ref{sec:2.2}.

\begin{figure}[!t]
  \centering
  \psfrag{a}[c][c][0.65]{BS $1$}
  \psfrag{b}[c][c][0.65]{BS $K$}
  \psfrag{c}[c][c][0.65]{relay $1$}
  \psfrag{d}[c][c][0.65]{relay $R$}
  \psfrag{e}[c][c][0.65]{MS $1$}
  \psfrag{f}[c][c][0.65]{MS $M$}
  \psfrag{g}[c][c][0.65]{MS}
  \psfrag{i}[c][c][0.65]{$(K-1)M+1$}
  \psfrag{h}[c][c][0.65]{MS $K\!M$}
  \psfrag{ha}[c][c][0.7]{$\mathbf{H}_{\mathrm{RB}}^{\left(1,1\right)}$}
  \psfrag{hb}[c][c][0.7]{$\mathbf{H}_{\mathrm{RB}}^{\left(R,1\right)}$}
  \psfrag{hc}[c][c][0.7]{$\mathbf{H}_{\mathrm{RB}}^{\left(1,K\right)}$}
  \psfrag{hd}[c][c][0.7]{$\mathbf{H}_{\mathrm{RB}}^{\left(R,K\right)}$}
  \psfrag{h1}[c][c][0.7]{$\mathbf{H}_{\mathrm{MR}}^{\left(1,1\right)}$}
  \psfrag{h2}[c][c][0.7]{$\mathbf{H}_{\mathrm{MR}}^{\left(M,1\right)}$}
  \psfrag{h3}[c][c][0.7]{$\mathbf{H}_{\mathrm{MR}}^{\left((K-1)M+1,1\right)}$}
  \psfrag{h4}[c][c][0.7]{$\mathbf{H}_{\mathrm{MR}}^{\left(M,R\right)}$}
  \psfrag{h5}[c][c][0.7]{$\mathbf{H}_{\mathrm{MR}}^{\left(K\!M,1\right)}$}
  \psfrag{h6}[c][c][0.7]{$\mathbf{H}_{\mathrm{MR}}^{\left(1,R\right)}$}
  \psfrag{h8}[c][c][0.7]{$\mathbf{H}_{\mathrm{MR}}^{\left((K-1)M+1,R\right)}$}
  \psfrag{h9}[c][c][0.7]{$\mathbf{H}_{\mathrm{MR}}^{\left(K\!M,R\right)}$}
  \includegraphics[width=3.5in]{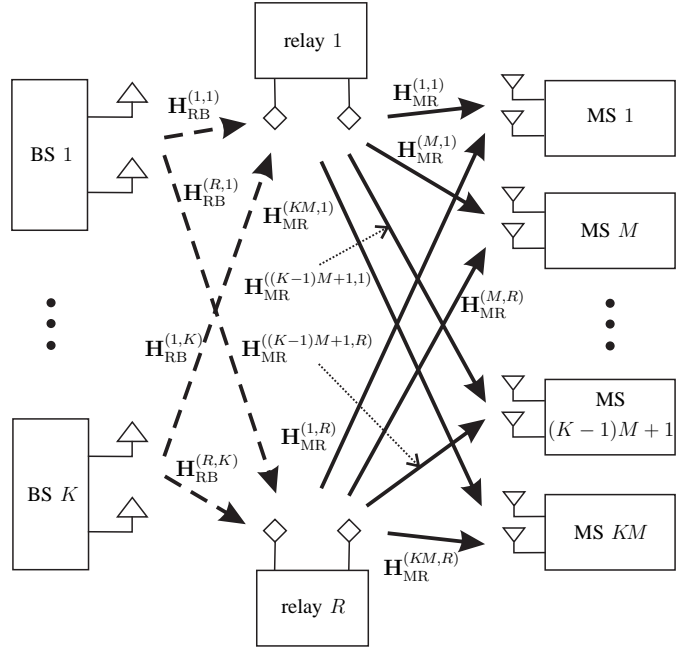}\\
  \caption{A $K$ cell scenario with $K$ BSs, $R$ relays and $KM$ MSs. The transmissions in the first and the second time slot are illustrated by the dotted and solid arrows, respectively.}\label{fig:sce_1}
\end{figure}

A downlink cellular scenario consisting of $K$ cells is considered. Each cell contains a BS with $N_{\mathrm{B}}$ antennas, and $M$ MSs with $N_{\mathrm{M}}$ antennas each. We first assume that the direct channels between the BSs and the MSs are relatively weak due to the radio environment so that they can be neglected. To enable the communication between the BSs and the MSs, $R$ half-duplex relays with $N_{\mathrm{R}}$ antennas each are deployed in the scenario. The transmission takes place in two subsequent time slots as illustrated in Fig. \ref{fig:sce_1}. In the first time slot, the BSs transmit to the relays. In the second time slot, the relays retransmit a linearly processed version of what they received in the first time slot to the MSs. The channels between the communication parties are assumed to remain constant during the transmission. To simplify the discussions, we assume that each MS receives a single desired data symbol from the corresponding BS. Accordingly, each BS transmits simultaneously 
$M$ complex valued data symbols with $M\leq N_{\mathrm{B}}$. A direct extension to the case where multiple data symbols are desired will be briefly discussed in Section \ref{sec:7}.

Let $k \in \{1,\ldots,K\}$, $m \in \{1,\ldots,KM\}$, and $r \in \{1,\ldots,R\}$ denote the indices of the BSs, the MSs, and the relays, respectively. Then, the data symbol transmitted by the corresponding BS for the $m$-th MS is denoted by $d^{\left(m\right)}\in \mathbb{C}$ and all the data symbols transmitted by the $k$-th BS are denoted by the vector $\mathbf{d}^{\left(k\right)} \in \mathbb{C}^M$. For each BS $k$, the transmitted data symbols are pre-processed by a linear transmit filter denoted by $\mathbf{V}^{\left(k\right)}\in \mathbb{C}^{N_{\mathrm{B}}\times M}$. The signal vector transmitted by the $k$-th BS reads
\begin{equation}\label{eq:tran_BS}
    \mathbf{s}_{\mathrm{B}}^{\left(k\right)}=\mathbf{V}^{\left(k\right)}\mathbf{d}^{\left(k\right)}.
\end{equation}
The received signal vector at the $r$-th relay is
\begin{equation}\label{eq:rece_relay}
    \mathbf{e}_{\mathrm{R}}^{\left(r\right)}=\sum\limits_{k=1}^{K}\mathbf{H}_{\mathrm{RB}}^{\left(r,k\right)}\mathbf{s}_{\mathrm{B}}^{\left(k\right)}+\mathbf{n}_{\mathrm{R}}^{\left(r\right)},
\end{equation}
where $\mathbf{H}_{\mathrm{RB}}^{\left(r,k\right)}\in \mathbb{C}^{N_{\mathrm{R}}\times N_{\mathrm{B}}}$ denotes the channel matrix between the $k$-th BS and the $r$-th relay, and $\mathbf{n}_{\mathrm{R}}^{\left(r\right)} \in \mathbb{C}^{N_{\mathrm{R}}\times 1}$ represents the noises at the different antennas of the relay, which are assumed to be independently identically distributed (i.i.d.) Gaussian with zero mean and variance $\sigma^{2}$. It is assumed that the number $N_{\mathrm{R}}$ of antennas at a relay is not large enough to spatially separate the received signals, i.e., $N_{\mathrm{R}}<KM$. Therefore, the amplify and forward relaying protocol is considered. The $r$-th relay linearly processes its received signals with the matrix $\mathbf{G}^{\left(r\right)}\in \mathbb{C}^{N_{\mathrm{R}}\times N_{\mathrm{R}}}$ and the transmitted signal of the $r$-th relay is denoted by
\begin{equation}\label{eq:tran_relay}
    \mathbf{s}_{\mathrm{R}}^{\left(r\right)}=\mathbf{G}^{\left(r\right)}\mathbf{e}_{\mathrm{R}}^{\left(r\right)}.
\end{equation}
Furthermore, the received signal vector at the $m$-th MS is
\begin{equation}\label{eq:rec_y_1}
    \mathbf{e}_{\mathrm{M}}^{\left(m\right)}=\sum\limits_{r=1}^{R}\mathbf{H}_{\mathrm{MR}}^{\left(m,r\right)}
    \mathbf{s}_{\mathrm{R}}^{\left(r\right)}+\mathbf{n}_{\mathrm{M}}^{\left(m\right)},
\end{equation}
where $\mathbf{H}_{\mathrm{MR}}^{\left(m,r\right)}\in \mathbb{C}^{N_{\mathrm{M}}\times N_{\mathrm{R}}}$ denotes the channel matrix between the $r$-th relay and the $m$-th MS, and $\mathbf{n}_{\mathrm{M}}^{\left(m\right)}\in \mathbb{C}^{N_{\mathrm{M}}\times 1}$ represents the noises at the MS, which are also assumed to be i.i.d. Gaussian with zero mean and variance $\sigma^{2}$. Then the $m$-th MS can linearly post-process its received signal vector $\mathbf{e}_{\mathrm{M}}^{\left(m\right)}$ using a linear receive filter $\mathbf{u}^{\left(m\right)}\in \mathbb{C}^{N_{\mathrm{M}}\times 1}$ to obtain the estimated data symbol as
\begin{align}\label{eq:rec_d_hat_1}
    \hat{d}^{\left(m\right)} &=\mathbf{u}^{\left(m\right)\ast \mathrm{T}}\mathbf{e}_{\mathrm{M}}^{\left(m\right)} \nonumber \\
    &=\mathbf{u}^{\left(m\right)\ast \mathrm{T}}\Big(\sum\limits_{r=1}^{R}\sum\limits_{k=1}^{K}\mathbf{H}_{\mathrm{MR}}^{\left(m,r\right)}
    \mathbf{G}^{\left(r\right)}\mathbf{H}_{\mathrm{RB}}^{\left(r,k\right)}
    \mathbf{V}^{\left(k\right)}\mathbf{d}^{\left(k\right)} \nonumber \\
    &+\sum\limits_{r=1}^{R}\mathbf{H}_{\mathrm{MR}}^{\left(m,r\right)}
    \mathbf{G}^{\left(r\right)}\mathbf{n}_{\mathrm{R}}^{\left(r\right)}
    +\mathbf{n}_{\mathrm{M}}^{\left(m\right)}\Big).
\end{align}

Suppose the duration of each time slot is normalized to one, it is assumed that the transmitted data symbols are uncorrelated and they have the same average power
\begin{equation}\label{eq:avg_egy}
    \E{\left|d^{\left(m\right)}\right|^{2}}=P_{\mathrm{d}},
\end{equation}
for all $m=1,\ldots,KM$ where $\E{\cdot}$ denotes the expectation. Moreover, the sum power constraint at the BSs is given by
\begin{equation}\label{eq:tot_BS}
    \sum\limits_{k=1}^{K} \tr{\E{\mathbf{s}_{\mathrm{B}}^{\left(k\right)}\mathbf{s}_{\mathrm{B}}^{\left(k\right)\ast \mathrm{T}}}}\leq P_{\mathrm{B}}.
\end{equation}
The sum power constraint at the relays is given by
\begin{equation}\label{eq:tot_relay}
    \sum\limits_{r=1}^{R} \tr{\E{\mathbf{s}_{\mathrm{R}}^{\left(r\right)}\mathbf{s}_{\mathrm{R}}^{\left(r\right)\ast \mathrm{T}}}}\leq P_{\mathrm{R}}.
\end{equation}

\subsection{Single-Hop Interference Broadcast Scenario} \label{sec:2.2}
\begin{figure}[!t]
  \centering
  \psfrag{a}[c][c][0.65]{BS $1$}
  \psfrag{b}[c][c][0.65]{BS $K$}
  \psfrag{e}[c][c][0.65]{MS $1$}
  \psfrag{f}[c][c][0.65]{MS $M$}
  \psfrag{g}[c][c][0.65]{MS}
  \psfrag{i}[c][c][0.65]{$(K-1)M+1$}
  \psfrag{h}[c][c][0.65]{MS $K\!M$}
  \psfrag{h1}[c][c][0.7]{$\mathbf{H}_{\mathrm{MB}}^{\left(1,1\right)}$}
  \psfrag{h2}[c][c][0.7]{$\mathbf{H}_{\mathrm{MB}}^{\left(M,1\right)}$}
  \psfrag{h3}[c][c][0.7]{$\mathbf{H}_{\mathrm{MB}}^{\left((K-1)M+1,1\right)}$}
  \psfrag{h4}[c][c][0.7]{$\mathbf{H}_{\mathrm{MB}}^{\left(K\!M,1\right)}$}
  \psfrag{h6}[c][c][0.7]{$\mathbf{H}_{\mathrm{MB}}^{\left(1,K\right)}$}
  \psfrag{h7}[c][c][0.7]{$\mathbf{H}_{\mathrm{MB}}^{\left(M,K\right)}$}
  \psfrag{h8}[c][c][0.7]{$\mathbf{H}_{\mathrm{MB}}^{\left((K-1)M+1,K\right)}$}
  \psfrag{h9}[c][c][0.7]{$\mathbf{H}_{\mathrm{MB}}^{\left(K\!M,K\right)}$}
  \includegraphics[width=3.5in]{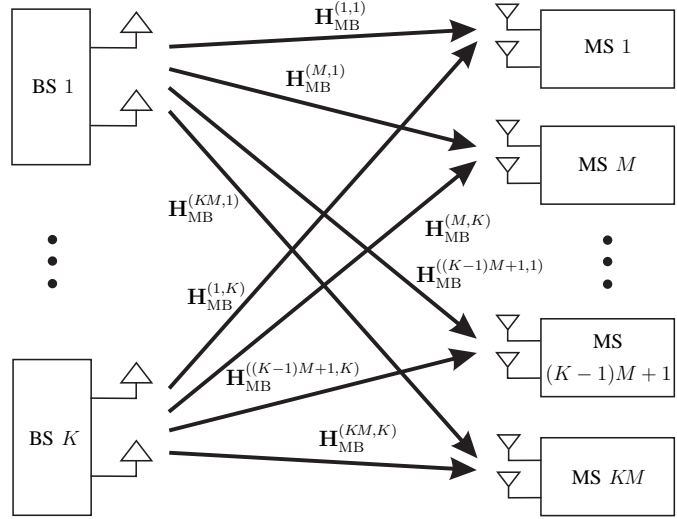}
  \caption{A $K$ cell scenario with $K$ BSs and $KM$ MSs.}\label{fig:sce_2}
\end{figure}

The second scenario we consider is similar to the one described in Section \ref{sec:2.1}, except that the direct channels between the BSs and MSs are assumed to be usable and no relays are deployed. The MSs receive signals directly from the BSs within a single time slot as illustrated in Fig. \ref{fig:sce_2}. Therefore, the received signal vector at the $m$-th MS reads
\begin{equation}\label{eq:rec_direct}
    \mathbf{e}_{\mathrm{M}}^{\left(m\right)}=\sum\limits_{k=1}^{K}\mathbf{H}_{\mathrm{MB}}^{\left(m,k\right)}
    \mathbf{s}_{\mathrm{B}}^{\left(k\right)}+\mathbf{n}_{\mathrm{M}}^{\left(m\right)},
\end{equation}
where $\mathbf{H}_{\mathrm{MB}}^{\left(m,k\right)}\in \mathbb{C}^{N_{\mathrm{M}}\times N_{\mathrm{B}}}$ denotes the channel matrix between the $m$-th MS and the $k$-th BS. Similar to (\ref{eq:rec_d_hat_1}), the estimated data symbol at the $m$-th MS is calculated as
\begin{equation}\label{eq:dhat_direct}
    \hat{d}^{\left(m\right)}=\mathbf{u}^{\left(m\right)\ast \mathrm{T}}\left(\sum\limits_{k=1}^{K}\mathbf{H}_{\mathrm{MB}}^{\left(m,k\right)}\mathbf{V}^{\left(k\right)}\mathbf{d}^{\left(k\right)}
    +\mathbf{n}_{\mathrm{M}}^{\left(m\right)}\right).
\end{equation}
Furthermore, only the power constraint (\ref{eq:tot_BS}) at the BSs is relevant for the single-hop scenario.

\section{Two-hop transmission scheme}
\label{sec:5}
\subsection{Problem formulation}
\label{sec:5.1}
To simplify the notations for the rest of this paper, we will partition the system variables into three disjoint sets with the variables being kept in a certain order for plugging them in a function argument, namely the tuple of the transmit filters
\begin{equation}\label{eq:tf_set}
    \mathbb{V}=\left(\mathbf{V}^{\left(1\right)},\ldots,\mathbf{V}^{\left(K\right)}\right),
\end{equation}
the tuple of the relay processing matrices
\begin{equation}\label{eq:rpm_set}
    \mathbb{G}=\left(\mathbf{G}^{\left(1\right)},\ldots,\mathbf{G}^{\left(R\right)}\right),
\end{equation}
and the tuple of the receive filters
\begin{equation}\label{eq:rpm_set}
    \mathbb{U}=\left(\mathbf{u}^{\left(1\right)},\ldots,\mathbf{u}^{\left(KM\right)}\right).
\end{equation}

In this section, we formulate the sum rate maximization problem for the two-hop interference broadcast scenario described in Section \ref{sec:2.1}. Using the notations introduced above, it can be observed from (\ref{eq:rec_d_hat_1}) that the estimated data symbol $\hat{d}^{\left(m\right)}$ at the $m$-th MS is a tri-affine function of the tuple $\mathbb{V}$ of the transmit filters, the tuple $\mathbb{G}$ of the relay processing matrices, and the receive filter $\mathbf{u}^{\left(m\right)}$. Let $\mathbf{v}^{\left(k,m\right)}$ denote the $m$-th column of $\mathbf{V}^{\left(k\right)}$. Then, (\ref{eq:rec_d_hat_1}) can be rewritten as
\begin{align}\label{eq:rec_MS2}
    \hat{d}^{\left(m\right)}=
    \mathbf{u}^{\left(m\right)\ast\mathrm{T}}\left(\mathbf{q}^{\left(m\right)}d^{\left(m\right)}+\mathbf{z}^{\left(m\right)}\right),
\end{align}
where
\begin{equation}\label{eq:eff_ch}
    \mathbf{q}^{\left(m\right)}=\sum\limits_{r=1}^{R}\mathbf{H}_{\mathrm{MR}}^{\left(m,r\right)}
    \mathbf{G}^{\left(r\right)}\mathbf{H}_{\mathrm{RB}}^{\left(r,k\right)}\mathbf{v}^{\left(k,m\right)}
\end{equation}
is the effective useful link of the $m$-th MS including the relays and the transmit filter vector $\mathbf{v}^{\left(k,m\right)}$. Let $\mathbf{\Upsilon}^{\left(m\right)}$ be an $M\times M$ diagonal matrix where all the diagonal elements are ones except for the $m$-th diagonal element being zero. The received interference plus noise at the $m$-th MS is given by
\begin{align}\label{eq:interference_noise}
    \mathbf{z}^{\left(m\right)}&=\sum\limits_{r=1}^{R}\mathbf{H}_{\mathrm{MR}}^{\left(m,r\right)}
    \mathbf{G}^{\left(r\right)}\mathbf{H}_{\mathrm{RB}}^{\left(r,k\right)}
    \mathbf{V}^{\left(k\right)}\mathbf{\Upsilon}^{\left(m\right)}\mathbf{d}^{\left(k\right)} \nonumber \\
    &+\sum\limits_{r=1}^{R}\mathbf{H}_{\mathrm{MR}}^{\left(m,r\right)}
    \mathbf{G}^{\left(r\right)}\sum\limits_{l\neq k}\mathbf{H}_{\mathrm{RB}}^{\left(r,l\right)}
    \mathbf{V}^{\left(l\right)}\mathbf{d}^{\left(l\right)} \nonumber \\
    &+\sum\limits_{r=1}^{R}\mathbf{H}_{\mathrm{MR}}^{\left(m,r\right)}
    \mathbf{G}^{\left(r\right)}\mathbf{n}_{\mathrm{R}}^{\left(r\right)}
    +\mathbf{n}_{\mathrm{M}}^{\left(m\right)}.
\end{align}
The first term and the second term of (\ref{eq:interference_noise}) represent the received intra-cell and inter-cell interference, respectively. The last two terms of (\ref{eq:interference_noise}) describe the noises received by the $m$-th MS, including the noise retransmitted by the relays. Based on this, the receive SINR at the $m$-th MS can be written as
\begin{equation}\label{eq:SINR}
    \gamma^{\left(m\right)}=\frac{P_{\mathrm{d}}\left|\mathbf{u}^{\left(m\right)\ast \mathrm{T}}\mathbf{q}^{\left(m\right)}\right|^{2}}{\E{\left|\mathbf{u}^{\left(m\right)\ast \mathrm{T}}\mathbf{z}^{\left(m\right)}\right|^{2}}},
\end{equation}
and thus, the sum rate is calculated as
\begin{equation}\label{eq:sumRate}
    \mathrm{C}\left(\mathbb{V},\mathbb{G},\mathbb{U}\right)=\sum\limits_{m=1}^{KM}\log{1+\gamma^{\left(m\right)}},
\end{equation}
which is a function of the tuples of variables $\mathbb{V}$, $\mathbb{G}$, and $\mathbb{U}$. For the considered two-hop transmission scheme, the sum rate maximization problem for optimizing the transmit filters, the relay processing matrices and the receive filters with the sum power constraints at the BSs and at the relays can be stated as
\begin{equation}\label{eq:obj1}
    \left(\mathbb{V}_{\mathrm{opt}},\mathbb{G}_{\mathrm{opt}},\mathbb{U}_{\mathrm{opt}}\right)=\argmax_{\left(\mathbb{V},
    \mathbb{G},\mathbb{U}\right)}\left\{\mathrm{C}\left(\mathbb{V},\mathbb{G},\mathbb{U}\right)\right\}
\end{equation}
subject to
\begin{equation}\label{eq:con1.1}
    P_{\mathrm{d}}\sum\limits_{k=1}^{K}\tr{\mathbf{V}^{\left(k\right)}\mathbf{V}^{\left(k\right)\ast \mathrm{T}}}\leq P_{\mathrm{B}}
\end{equation}
and
\begin{align}\label{eq:con1.2}
    &P_{\mathrm{d}}\sum\limits_{r=1}^{R}\tr{\mathbf{G}^{\left(r\right)}\sum\limits_{k=1}^{K}\mathbf{H}_{\mathrm{RB}}^{\left(r,k\right)}
    \mathbf{V}^{\left(k\right)}\mathbf{V}^{\left(k\right)\ast \mathrm{T}}\mathbf{H}_{\mathrm{RB}}^{\left(r,k\right)\ast \mathrm{T}}\mathbf{G}^{\left(r\right)\ast \mathrm{T}}} \nonumber \\
    &+\sigma^{2}\sum\limits_{r=1}^{R}\tr{\mathbf{G}^{\left(r\right)}\mathbf{G}^{\left(r\right)\ast \mathrm{T}}}\leq P_{\mathrm{R}},
\end{align}
where the constraints of (\ref{eq:con1.1}) and (\ref{eq:con1.2}) follow from (\ref{eq:tot_BS}) and (\ref{eq:tot_relay}), respectively. The sum power constraint of (\ref{eq:con1.1}) at the BSs is a convex set of the transmit filters. Furthermore, the sum power constraint of (\ref{eq:con1.2}) at the relays is a biconvex set of the transmit filters and the relay processing matrices. However, the objective function -- the sum rate function -- is not a concave function of $\mathbb{V}$, $\mathbb{G}$, and $\mathbb{U}$. Therefore, the optimization problem of (\ref{eq:obj1})--(\ref{eq:con1.2}) is a non-convex problem.

\subsection{Signal to Interference plus Noise Ratio}
\label{sec:5.2}
With a closer look at the structure of the sum rate function of (\ref{eq:sumRate}), one can observe that the achieved rate at a MS is a logarithmic function of $1+$SINR. Basically, the main difficulty of handling the SINR function of (\ref{eq:SINR}) is that both its nominator and denominator are functions of the system variables, see (\ref{eq:eff_ch}) and (\ref{eq:interference_noise}). In order to reformulate the optimization problem of (\ref{eq:obj1})--(\ref{eq:con1.2}) as a multi-convex optimization problem, a term related to the SINR is introduced in the following proposition.
\begin{mydef1}\label{def:1}
Let $w^{\left(m\right)}\in \mathbb{C}$ be a scaling factor which scales the $m$-th transmitted data symbol $d^{\left(m\right)}$. Then, the function
\begin{equation}\label{eq:eta_1}
    \eta^{\left(m\right)}\left( w^{\left(m\right)} \right) =\frac{\E{\left|w^{\left(m\right)}d^{\left(m\right)}\right|^{2}}}
    {\E{\left|\hat{d}^{\left(m\right)}-w^{\left(m\right)}d^{\left(m\right)}\right|^{2}}}
\end{equation}
has a single maximum being equal to $1+\gamma^{\left(m\right)}$, where $\gamma^{\left(m\right)}$ is defined in \eqref{eq:SINR}.
\end{mydef1}
\begin{proof}
Using the function
\begin{align}\label{eq:g_fun}
    \mathrm{g}&\left(\mathbb{V},\mathbb{G},\mathbf{u}^{\left(m\right)},w^{\left(m\right)}\right)=
    \E{\left|\hat{d}^{\left(m\right)}-w^{\left(m\right)}d^{\left(m\right)}\right|^{2}} \nonumber \\
    &=P_{\mathrm{d}}\Big|\mathbf{u}^{\left(m\right)\ast \mathrm{T}}\mathbf{q}^{\left(m\right)}-w^{\left(m\right)}\Big|^{2}+\E{\left|\mathbf{u}^{\left(m\right)\ast \mathrm{T}}\mathbf{z}^{\left(m\right)}\right|^{2}},
\end{align}
(\ref{eq:eta_1}) can be written as
\begin{equation}\label{eq:eta_2}
    \eta^{\left(m\right)}\left( w^{\left(m\right)} \right)=\frac{P_{\mathrm{d}}\left|w^{\left(m\right)}\right|^{2}}
    {\mathrm{g}\left(\mathbb{V},{G},\mathbf{u}^{\left(m\right)},w^{\left(m\right)}\right)}.
\end{equation}
Since $\hat{d}^{\left(m\right)}$ described in (\ref{eq:rec_d_hat_1}) is a tri-affine function of $\mathbb{V}$, $\mathbb{G}$, and $\mathbf{u}^{\left(m\right)}$, the function \\  $\mathrm{g}\left(\mathbb{V},\mathbb{G},\mathbf{u}^{\left(m\right)},w^{\left(m\right)}\right)$ described in (\ref{eq:g_fun}) is a tri-convex function of $\mathbb{V}$, $\mathbb{G}$ and $\mathbf{u}^{\left(m\right)}$ for a fixed $w^{\left(m\right)}$. By calculating the general derivative of $\eta^{\left(m\right)}$ with respect to $w^{\left(m\right)}$ and setting the result to zero, two stationary points can be calculated as
\begin{equation}\label{eq:w_0}
  w^{\left(m\right)}_0=0
\end{equation}
and
\begin{equation}\label{eq:w_opt}
    w^{\left(m\right)}_{\mathrm{opt}}=
    \frac{P_{\mathrm{d}} \left|\mathbf{u}^{\left(m\right)\ast\mathrm{T}}
    \mathbf{q}^{\left(m\right)}\right|^{2}+\E{\left|\mathbf{u}^{\left(m\right)\ast\mathrm{T}}\mathbf{z}^{\left(m\right)}\right|^{2}}}{P_{\mathrm{d}}\mathbf{q}^{\left(m\right)\ast \mathrm{T}}\mathbf{u}^{\left(m\right)}}.
\end{equation}
By substituting (\ref{eq:w_0}) and (\ref{eq:w_opt}) in (\ref{eq:eta_2}), the values of $\eta^{\left(m\right)}$ at $w^{\left(m\right)}_0$ and $w^{\left(m\right)}_{\mathrm{opt}}$, respectively, are calculated as
\begin{equation}
  \eta^{\left(m\right)}(w^{\left(m\right)}_0)=0
\end{equation}
and
\begin{align}\label{eq:eta_opt}
    \eta^{\left(m\right)}(w^{\left(m\right)}_{\mathrm{opt}})&=\frac{P_{\mathrm{d}}\left|\mathbf{u}^{\left(m\right)\ast \mathrm{T}}\mathbf{q}^{\left(m\right)}\right|^{2}+\E{\left|\mathbf{u}^{\left(m\right)\ast \mathrm{T}}\mathbf{z}^{\left(m\right)}\right|^{2}}}{\E{\left|\mathbf{u}^{\left(m\right)\ast \mathrm{T}}\mathbf{z}^{\left(m\right)}\right|^{2}}} \nonumber \\
    &= 1+\gamma^{\left(m\right)}.
\end{align}
Considering the fact that the function $\eta^{\left(m\right)}$ described in (\ref{eq:eta_1}) is non-negative and
\begin{equation}\label{eq:endpoint_min}
    \lim_{|w^{\left(m\right)}| \rightarrow \infty}\eta^{\left(m\right)}=1,
\end{equation}
the function $\eta^{\left(m\right)}$ must achieve its maximum at $w^{\left(m\right)}_{\mathrm{opt}}$.
\end{proof}

The nice property of $\eta^{\left(m\right)}$ is that just its denominator is a function of the system variables $\mathbb{V}$, $\mathbb{G}$, and $\mathbf{u}^{\left(m\right)}$ whereas for $\gamma^{\left(m\right)}$ defined in (\ref{eq:SINR}), both the nominator and the denominator are functions of the system variables.
\subsection{Problem reformulation}
\label{sec:5.3}
In the previous section, it has been shown that the term $\eta^{\left(m\right)}$ is equivalent to the received SINR at the $m$-th MS when the scaling factor $w^{\left(m\right)}$ is optimized using (\ref{eq:w_opt}). Let
\begin{equation}\label{eq:rpm_set}
    \mathbf{w}=\left(w^{\left(1\right)},\ldots,w^{\left(KM\right)}\right)^\textrm{T}
\end{equation}
be a vector of the scaling factors and let the elements of $\mathbf{w}_{\mathrm{opt}}$ be chosen as (\ref{eq:w_opt}). Then, the function
\begin{equation}\label{eq:f1}
    \mathrm{f}_{\mathrm{2hop}}\left(\mathbb{V},\mathbb{G},\mathbb{U},\mathbf{w}\right)=\sum\limits_{k=1}^{K}\sum\limits_{m=1}^{M}\log{\eta^{\left(m\right)}}
\end{equation}
is equivalent to the sum rate function of (\ref{eq:sumRate}) in the sense that both have the same local and global maxima if $\mathbf{w}=\mathbf{w}_{\mathrm{opt}}$ holds.
To show the concavity of $\mathrm{f}_{\mathrm{2hop}}$ with respect to the tuples $\mathbb{V}$, $\mathbb{G}$ and $\mathbb{U}$, using (\ref{eq:eta_2}), (\ref{eq:f1}) can be rewritten as
\begin{align}\label{eq:obj2.1}
    \mathrm{f}_{\mathrm{2hop}}\left(\mathbb{V},\mathbb{G},\mathbb{U},\mathbf{w}\right)=&\sum\limits_{k=1}^{K}\sum\limits_{m=1}^{M}\log{P_{\mathrm{d}}\left|w^{\left(m\right)}\right|^{2}} \nonumber\\
    &-\sum\limits_{k=1}^{K}\sum\limits_{m=1}^{M}\log{\mathrm{g}\left(\mathbb{V},\mathbb{G},\mathbf{u}^{\left(m\right)},w^{\left(m\right)}\right)}.
\end{align}
In (\ref{eq:obj2.1}), just the second term includes the system variables. Although the function \\$\mathrm{g}\left(\mathbb{V},\mathbb{G},\mathbf{u}^{\left(m\right)},w^{\left(m\right)}\right)$ is a tri-convex function of $\mathbb{V}$, $\mathbb{G}$ and $\mathbf{u}^{\left(m\right)}$ when $w^{\left(m\right)}$ is fixed, \\$\log{\mathrm{g}\left(\mathbb{V},\mathbb{G},\mathbf{u}^{\left(m\right)},w^{\left(m\right)}\right)}$ is not necessarily convex \cite{C78}. Accordingly, we aim at finding a new equivalent objective function which is linear in $\mathrm{g}\left(\mathbb{V},\mathbb{G},\mathbf{u}^{\left(m\right)},w^{\left(m\right)}\right)$ such that we can exploit the fact that $\mathrm{g}\left(\mathbb{V},\mathbb{G},\mathbf{u}^{\left(m\right)},w^{\left(m\right)}\right)$ is a multi-convex function of the system variables.
\subsection{Multi-convex problem formulation}
\label{sec:5.4}
In this section, the optimization problem of (\ref{eq:obj1})--(\ref{eq:con1.2}) is reformulated as a multi-convex optimization problem. Let
\begin{equation}\label{eq:rpm2_set}
    \mathbf{t}=\left(t^{\left(1\right)},\ldots,t^{\left(KM\right)}\right)^\text{T}
\end{equation}
be a vector of additional scaling factors. Then, the function
\begin{align}\label{eq:obj2.2}
    &\mathrm{b}_{\mathrm{2hop}}\left(\mathbb{V},\mathbb{G},\mathbb{U},\mathbf{w},\mathbf{t}\right)=
    \sum\limits_{k=1}^{K}\sum\limits_{m=1}^{M}\bigg(\log{P_{\mathrm{d}}\left|w^{\left(m\right)}\right|^{2}}\nonumber\\
    &+\log{t^{\left(m\right)}}
    -\frac{t^{\left(m\right)}}{\ln{2}}\mathrm{g}\left(\mathbb{V},\mathbb{G},\mathbf{u}^{\left(m\right)},w^{\left(m\right)}\right)\bigg)
\end{align}
is obviously a concave function of $\mathbf{t}$. By taking the first order derivative of $\mathrm{b}_{\mathrm{2hop}}$ with respect to $t^{\left(m\right)}$ and setting the result to zero, the optimum scaling factor $t^{\left(m\right)}$ is calculated as
\begin{equation}\label{eq:t_opt}
    t_{\mathrm{opt}}^{\left(m\right)}=\frac{1}{\mathrm{g}\left(\mathbb{V},\mathbb{G},\mathbf{u}^{\left(m\right)},w^{\left(m\right)}\right)}.
\end{equation}
Substituting (\ref{eq:t_opt}) in (\ref{eq:obj2.2}) yields
\begin{equation}\label{eq:obj_equi}
    \mathrm{b}_{\mathrm{2hop}}\left(\mathbb{V},\mathbb{G},\mathbb{U},\mathbf{w},\mathbf{t}_{\mathrm{opt}}\right)
    =\mathrm{f}_{\mathrm{2hop}}\left(\mathbb{V},\mathbb{G},\mathbb{U},\mathbf{w}\right)-\frac{KM}{\ln{2}}.
\end{equation}
From (\ref{eq:obj_equi}), it can be concluded that the new objective function $\mathrm{b}_{\mathrm{2hop}}\left(\mathbb{V},\mathbb{G},\mathbb{U},\mathbf{w},\mathbf{t}\right)$ is equivalent to the sum rate function in the sense that they both have the same global and local maxima if the optimum scaling factors in $\mathbf{w}$ and $\mathbf{t}$ are chosen. Moreover, the function $\mathrm{b}_{\mathrm{2hop}}\left(\mathbb{V},\mathbb{G},\mathbb{U},\mathbf{w},\mathbf{t}\right)$ has a single maximum at $\mathbf{w}=\mathbf{w}_{\mathrm{opt}}$ if $\mathbb{V}$, $\mathbb{G}$, $\mathbb{U}$, and $\mathbf{t}$ are fixed, and the function $\mathrm{b}_{\mathrm{2hop}}\left(\mathbb{V},\mathbb{G},\mathbb{U},\mathbf{w}_{\mathrm{opt}},\mathbf{t}\right)$ is
\begin{itemize}
  \item a concave function of $\mathbf{t}$ if $\mathbb{V}$, $\mathbb{G}$, $\mathbb{U}$, and $\mathbf{w}$ are fixed because the logarithm is a concave monotonically increasing function,
  \item a concave function of $\mathbb{V}$ if $\mathbf{t}$, $\mathbb{G}$, $\mathbb{U}$, and $\mathbf{w}$ are fixed because $\mathrm{g}\left(\mathbb{V},\mathbb{G},\mathbf{u}^{\left(m\right)},w^{\left(m\right)}\right)$ is a convex function of $\mathbb{V}$,
  \item a concave function of $\mathbb{G}$ if $\mathbf{t}$, $\mathbb{V}$, $\mathbb{U}$, and $\mathbf{w}$ are fixed because $\mathrm{g}\left(\mathbb{V},\mathbb{G},\mathbf{u}^{\left(m\right)},w^{\left(m\right)}\right)$ is a convex function of $\mathbb{G}$, and
  \item a concave function of $\mathbb{U}$ if $\mathbf{t}$, $\mathbb{V}$, $\mathbb{G}$, and $\mathbf{w}$ are fixed because $\mathrm{g}\left(\mathbb{V},\mathbb{G},\mathbf{u}^{\left(m\right)},w^{\left(m\right)}\right)$, $\forall \: m$ is a convex function of $\mathbb{U}$.
\end{itemize}
Accordingly, the sum rate maximization problem of (\ref{eq:obj1})--(\ref{eq:con1.2}) can be equivalently formulated as a multi-convex optimization problem stated as
\begin{align}\label{eq:obj3}
    \big(\mathbb{V}_{\mathrm{opt}},\mathbb{G}_{\mathrm{opt}},&\mathbb{U}_{\mathrm{opt}},\mathbf{w}_{\mathrm{opt}},\mathbf{t}_{\mathrm{opt}}\big)\nonumber\\
    &=\argmax_{\left(\mathbb{V},\mathbb{G},\mathbb{U},\mathbf{w},\mathbf{t}\right)}
    \left\{\mathrm{b}_{\mathrm{2hop}}\left(\mathbb{V},\mathbb{G},\mathbb{U},\mathbf{w},\mathbf{t}\right)\right\}
\end{align}
subject to
\begin{equation}\label{eq:con3.1}
    P_{\mathrm{d}}\sum\limits_{k=1}^{K}\tr{\mathbf{V}^{\left(k\right)}\mathbf{V}^{\left(k\right)\ast \mathrm{T}}}\leq P_{\mathrm{B}}
\end{equation}
and
\begin{align}\label{eq:con3.2}
    &P_{\mathrm{d}}\sum\limits_{r=1}^{R}\tr{\mathbf{G}^{\left(r\right)}\sum\limits_{k=1}^{K}\mathbf{H}_{\mathrm{RB}}^{\left(r,k\right)}
    \mathbf{V}^{\left(k\right)}\mathbf{V}^{\left(k\right)\ast \mathrm{T}}\mathbf{H}_{\mathrm{RB}}^{\left(r,k\right)\ast \mathrm{T}}\mathbf{G}^{\left(r\right)\ast \mathrm{T}}} \nonumber \\
    &+\sigma^{2}\sum\limits_{r=1}^{R}\tr{\mathbf{G}^{\left(r\right)}\mathbf{G}^{\left(r\right)\ast \mathrm{T}}}\leq P_{\mathrm{R}}.
\end{align}
This problem is a multi-convex problem of $\mathbb{V}$, $\mathbb{G}$, $\mathbb{U}$, and $\mathbf{t}$.

The vectors $\mathbf{w}$ and $\mathbf{t}$ of the scaling factors can be optimized using (\ref{eq:w_opt}) and (\ref{eq:t_opt}), respectively. With fixed scaling factors, just the last term of (\ref{eq:obj2.2}) is relevant for optimizing the system variables and thus, the optimization problem (\ref{eq:obj3})--(\ref{eq:con3.2}) can be stated as
\begin{align}\label{eq:obj6}
    \big(\mathbb{V}_{\mathrm{min}}&,\mathbb{G}_{\mathrm{min}},\mathbb{U}_{\mathrm{min}}\big)\nonumber\\
    &=\argmin_{\left(\mathbb{V},\mathbb{G},\mathbb{U}\right)}
    \left\{\sum\limits_{k=1}^{K}\sum\limits_{m=1}^{M}\frac{t^{\left(m\right)}}{\ln{2}}
    \mathrm{g}\left(\mathbb{V},\mathbb{G},\mathbf{u}^{\left(m\right)},w^{\left(m\right)}\right)\right\}
\end{align}
subject to
\begin{equation}\label{eq:con6.1}
    P_{\mathrm{d}}\sum\limits_{k=1}^{K}\tr{\mathbf{V}^{\left(k\right)}\mathbf{V}^{\left(k\right)\ast \mathrm{T}}}\leq P_{\mathrm{B}}
\end{equation}
and
\begin{align}\label{eq:con6.2}
    &P_{\mathrm{d}}\sum\limits_{r=1}^{R}\tr{\mathbf{G}^{\left(r\right)}\sum\limits_{k=1}^{K}\mathbf{H}_{\mathrm{RB}}^{\left(r,k\right)}
    \mathbf{V}^{\left(k\right)}\mathbf{V}^{\left(k\right)\ast \mathrm{T}}\mathbf{H}_{\mathrm{RB}}^{\left(r,k\right)\ast \mathrm{T}}\mathbf{G}^{\left(r\right)\ast \mathrm{T}}} \nonumber \\
    &+\sigma^{2}\sum\limits_{r=1}^{R}\tr{\mathbf{G}^{\left(r\right)}\mathbf{G}^{\left(r\right)\ast \mathrm{T}}}\leq P_{\mathrm{R}}.
\end{align}
As described previously in Section \ref{sec:5.2}, the function $\mathrm{g}\left(\mathbb{V},\mathbb{G},\mathbf{u}^{\left(m\right)},w^{\left(m\right)}\right)$ is a tri-convex function of $\mathbb{V}$, $\mathbb{G}$, and $\mathbf{u}^{\left(m\right)}$ for fixed $w^{\left(m\right)}$. Moreover, the power constraints of (\ref{eq:con6.1}) and (\ref{eq:con6.2}) are a convex set and a biconvex set, respectively. Based on this, the optimization problem of (\ref{eq:obj6})--(\ref{eq:con6.2}) is a tri-convex problem for fixed $\mathbf{w}$ and $\mathbf{t}$. By taking the general derivative of $\mathrm{g}\left(\mathbb{V},\mathbb{G},\mathbf{u}^{\left(m\right)},w^{\left(m\right)}\right)$ with respect to $\mathbf{u}^{\left(m\right)}$ and setting the result to zero, the optimum receive filter is calculated as
\begin{align}\label{eq:u_opt}
    \mathbf{u}^{\left(m\right)}_{\mathrm{min}}=&\left(P_{\mathrm{d}}\mathbf{q}^{\left(m\right)}\mathbf{q}^{\left(m\right)\ast \mathrm{T}}+\E{\mathbf{z}^{\left(m\right)}\mathbf{z}^{\left(m\right)\ast \mathrm{T}}}\right)^{-1}\nonumber \\
    &\cdot P_{\mathrm{d}}w^{\left(m\right)\ast}\mathbf{q}^{\left(m\right)}.
\end{align}
The problem structure with respect to $\mathbb{V}$ with fixed $\mathbb{G}$, $\mathbb{U}$, $\mathbf{w}$, and $\mathbf{t}$, is a convex quadratically constrained quadratic problem \cite{huss13}. Tools from quadratic optimization can be applied to find the optimum transmit filters. Similarly, with fixed $\mathbb{V}$, $\mathbb{U}$, $\mathbf{w}$, and $\mathbf{t}$, the optimization problem (\ref{eq:obj6})--(\ref{eq:con6.2}) can be solved for the tuple $\mathbb{G}$ of the relay processing matrices using the conventional quadratic optimization tools.
\subsection{Iterative algorithm}
\label{sec:5.5}
In this section, an iterative algorithm which alternately maximizes the multi-concave objective function $\mathrm{b}_{\mathrm{2hop}}\left(\mathbb{V},\mathbb{G},\mathbb{U},\mathbf{w},\mathbf{t}\right)$ by sequentially optimizing $\mathbb{V}$, $\mathbb{G}$, $\mathbb{U}$, $\mathbf{w}$ and $\mathbf{t}$ is described. Let $\epsilon$ be an arbitrarily small tolerance value. Then, the proposed algorithm can be summarized as follows:
\begin{algorithmic}[1]
    \State set arbitrary initial values for $\mathbf{w}^{\left(0\right)}$, $\mathbf{t}^{\left(0\right)}$ and $\mathbb{U}^{\left(0\right)}$
    \State set feasible initial values for $\mathbb{V}^{\left(0\right)}$ and $\mathbb{G}^{\left(0\right)}$
    \Statex\Comment chosen such that the constraints of (\ref{eq:con3.1}) and (\ref{eq:con3.2}) hold
    \State in each iteration $i$
    \State \hspace{0.4cm}calculate $\mathbb{U}^{\left(i\right)}$ given $\mathbf{w}^{\left(i-1\right)}$, $\mathbb{V}^{\left(i-1\right)}$ and $\mathbb{G}^{\left(i-1\right)}$
    \Statex\Comment using (\ref{eq:u_opt})
    \State \hspace{0.4cm}calculate $\mathbb{V}^{\left(i\right)}$ given $\mathbf{w}^{\left(i-1\right)}$, $\mathbf{t}^{\left(i-1\right)}$, $\mathbb{U}^{\left(i\right)}$ and $\mathbb{G}^{\left(i-1\right)}$
    \Statex\Comment using quadratic optimization tools \cite{BV04}
    \State \hspace{0.4cm}calculate $\mathbb{G}^{\left(i\right)}$ given $\mathbf{w}^{\left(i-1\right)}$, $\mathbf{t}^{\left(i-1\right)}$, $\mathbb{U}^{\left(i\right)}$ and $\mathbb{V}^{\left(i\right)}$
    \Statex\Comment using quadratic optimization tools \cite{BV04}
    \State \hspace{0.4cm}calculate $\mathbf{t}^{\left(i\right)}$ given $\mathbf{w}^{\left(i-1\right)}$, $\mathbb{V}^{\left(i\right)}$, $\mathbb{G}^{\left(i\right)}$ and $\mathbb{U}^{\left(i\right)}$
    \Statex\Comment using (\ref{eq:t_opt})
    \State \hspace{0.4cm}calculate $\mathbf{w}^{\left(i\right)}$ given $\mathbb{V}^{\left(i\right)}$, $\mathbb{G}^{\left(i\right)}$ and $\mathbb{U}^{\left(i\right)}$ \Comment using (\ref{eq:w_opt})
    \State stop if $\big|\mathrm{b}_{\mathrm{2hop}}\left(\mathbb{V}^{\left(i\right)},\mathbb{G}^{\left(i\right)},\mathbb{U}^{\left(i\right)},\mathbf{w}^{\left(i\right)},\mathbf{t}^{\left(i\right)}\right)$
    \Statex $-\mathrm{b}_{\mathrm{2hop}}\left(\mathbb{V}^{\left(i-1\right)},\mathbb{G}^{\left(i-1\right)},\mathbb{U}^{\left(i-1\right)},\mathbf{w}^{\left(i-1\right)},\mathbf{t}^{\left(i-1\right)}\right)\big|\leq\epsilon$
\end{algorithmic}

From the multi-convex optimization literature \cite{GPK07} it is known that the above algorithm converges to a local maximum.
\section{Single-hop transmission scheme}
\label{sec:6}
\subsection{Problem formulation}
\label{sec:6.1}
In this section, we will show that the proposed multi-convex formulation of the sum rate and the iterative algorithm can be applied to the single-hop interference broadcast scenarios described in Section \ref{sec:2.2} as well.

From (\ref{eq:dhat_direct}), one can observe that $\hat{d}^{\left(m\right)}$ is a bi-affine function of the tuple $\mathbb{V}$ of transmit filters and the receive filter $\mathbf{u}^{\left(m\right)}$. Then (\ref{eq:dhat_direct}) can be rewritten as
\begin{align}\label{eq:rec_MS3}
    \hat{d}^{\left(m\right)}=
    \mathbf{u}^{\left(m\right)\ast\mathrm{T}}\left(\mathbf{q}^{\left(m\right)}d^{\left(m\right)}+\mathbf{z}^{\left(m\right)}\right),
\end{align}
where
\begin{equation}\label{eq:q_direct}
    \mathbf{q}^{\left(m\right)}=\mathbf{H}_{\mathrm{MB}}^{\left(m,k\right)}\mathbf{v}^{\left(k,m\right)}
\end{equation}
is the effective useful link corresponding to the $m$-th MS including the transmit filter vector $\mathbf{v}^{\left(k,m\right)}$, and the effective interference plus noise received at the $m$-th MS is
\begin{align}\label{eq:z_direct}
    \mathbf{z}^{\left(m\right)}&=\mathbf{H}_{\mathrm{MB}}^{\left(m,k\right)}\mathbf{V}^{\left(k\right)}\mathbf{\Upsilon}^{\left(m\right)}\mathbf{d}^{\left(k\right)}\nonumber\\
    &+\sum\limits_{l\neq k}\mathbf{H}_{\mathrm{MB}}^{\left(m,l\right)}\mathbf{V}^{\left(l\right)}\mathbf{d}^{\left(l\right)}+\mathbf{n}_{\mathrm{M}}^{\left(m\right)}.
\end{align}
In (\ref{eq:z_direct}), the first term and the second term represent the received intra-cell interference and the received inter-cell interference, respectively. The noise at the $m$-th MS is described by the last term of (\ref{eq:z_direct}).

Substituting (\ref{eq:q_direct}) and (\ref{eq:z_direct}) into (\ref{eq:SINR}), the receive SINR at the $m$-th MS can be calculated. Then, the sum rate can be calculated as
\begin{equation}\label{eq:sumRate2}
    \mathrm{C}\left(\mathbb{V},\mathbb{U}\right)=\sum\limits_{m=1}^{KM}\log{1+\gamma^{\left(m\right)}}
\end{equation}
and the sum rate maximization problem can be formulated as
\begin{equation}\label{eq:obj4}
    \left(\mathbb{V}_{\mathrm{opt}},\mathbb{U}_{\mathrm{opt}}\right)=\argmax_{\left(\mathbb{V},\mathbb{U}\right)}\left\{\mathrm{C}\left(\mathbb{V},\mathbb{U}\right)\right\}
\end{equation}
subject to
\begin{equation}\label{eq:con4}
    P_{\mathrm{d}}\sum\limits_{k=1}^{K}\tr{\mathbf{V}^{\left(k\right)}\mathbf{V}^{\left(k\right)\ast \mathrm{T}}}\leq P_{\mathrm{B}}.
\end{equation}
Similar to the two-hop scenario discussed in Section \ref{sec:5.1}, this problem is non-convex.

\subsection{Multi-convex problem formulation}
\label{sec:6.2}
For the single-hop transmission, the function $\mathrm{g}\left(\mathbb{V}, u^{\left(m\right)},w^{\left(m\right)}\right)$ described in (\ref{eq:g_fun}) can be redefined using (\ref{eq:rec_MS3}). The multi-concave objective function can be written as
\begin{align}\label{eq:obj5.2}
    &\mathrm{b}_{\mathrm{1hop}}\left(\mathbb{V},\mathbb{U},\mathbf{w},\mathbf{t}\right)=
    \sum\limits_{k=1}^{K}\sum\limits_{m=1}^{M}\bigg(\log{P_{\mathrm{d}}\left|w^{\left(m\right)}\right|^{2}}\nonumber\\
    &+\log{t^{\left(m\right)}}
    -\frac{t^{\left(m\right)}}{\ln{2}}\mathrm{g}\left(\mathbb{V},\mathbf{u}^{\left(m\right)},w^{\left(m\right)}\right)\bigg).
\end{align}
Based on this, the sum rate maximization problem can be formulated as a multi-convex optimization problem stated as
\begin{equation}\label{eq:obj5}
    \big(\mathbb{V}_{\mathrm{opt}},\mathbb{U}_{\mathrm{opt}},\mathbf{w}_{\mathrm{opt}},\mathbf{t}_{\mathrm{opt}}\big)
    =\argmax_{\left(V,U,W,T\right)}
    \left\{\mathrm{b}_{\mathrm{1hop}}\left(\mathbb{V},\mathbb{U},\mathbf{w},\mathbf{t}\right)\right\}
\end{equation}
subject to
\begin{equation}\label{eq:con5}
    P_{\mathrm{d}}\sum\limits_{k=1}^{K}\tr{\mathbf{V}^{\left(k\right)}\mathbf{V}^{\left(k\right)\ast \mathrm{T}}}\leq P_{\mathrm{B}}.
\end{equation}
This problem is a multi-convex optimization problem of $\mathbb{V}$ and $\mathbb{U}$ if the optimum $\mathbf{w}$ and $\mathbf{t}$ are a priori chosen. As described in Section \ref{sec:5.4}, the optimization problem of (\ref{eq:obj5})--(\ref{eq:con5}) can be solved alternatingly over $\mathbb{V}$, $\mathbb{U}$, $\mathbf{w}$, and $\mathbf{t}$. The iterative algorithm can be summarized as follows:
\begin{algorithmic}[1]
    \State set arbitrary initial values for $\mathbf{w}^{\left(0\right)}$, $\mathbf{t}^{\left(0\right)}$ and $\mathbb{U}^{\left(0\right)}$
    \State set feasible initial values for $\mathbb{V}^{\left(0\right)}$
    \Statex\Comment chosen such that the constraint of (\ref{eq:con5}) holds.
    \State in every iteration $i$
    \State \hspace{0.4cm}calculate $\mathbb{U}^{\left(i\right)}$ given $\mathbf{w}^{\left(i-1\right)}$ and $\mathbb{V}^{\left(i-1\right)}$\Comment using (\ref{eq:u_opt})
    \State \hspace{0.4cm}calculate $\mathbb{V}^{\left(i\right)}$ given $\mathbf{w}^{\left(i-1\right)}$, $\mathbf{t}^{\left(i-1\right)}$ and $\mathbb{U}^{\left(i\right)}$
    \Statex\Comment using quadratic optimization tools \cite{BV04}
    \State \hspace{0.4cm}calculate $\mathbf{t}^{\left(i\right)}$ given $\mathbf{w}^{\left(i-1\right)}$, $\mathbb{V}^{\left(i\right)}$ and $\mathbb{U}^{\left(i\right)}$\Comment using (\ref{eq:t_opt})
    \State \hspace{0.4cm}calculate $\mathbf{w}^{\left(i\right)}$ given $\mathbb{V}^{\left(i\right)}$ and $\mathbb{U}^{\left(i\right)}$\Comment using (\ref{eq:w_opt})
    \State stop if $\big|\mathrm{b}_{\mathrm{1hop}}\left(\mathbb{V}^{\left(i\right)},\mathbb{U}^{\left(i\right)},\mathbf{w}^{\left(i\right)},\mathbf{t}^{\left(i\right)}\right)$
    \Statex \hspace{0.9cm} $-\mathrm{b}_{\mathrm{1hop}}\left(\mathbb{V}^{\left(i-1\right)},\mathbb{U}^{\left(i-1\right)},\mathbf{w}^{\left(i-1\right)},\mathbf{t}^{\left(i-1\right)}\right)\big|\leq\epsilon$
\end{algorithmic}

\section{Further Discussions} \label{sec:disc}
The key idea of the algorithm proposed in this paper is to find a multi-concave objective function which is equivalent to the sum rate function in the sense that they have the same maxima, e.g., the function $\mathrm{b}_{\mathrm{2hop}}$ in (\ref{eq:obj2.2}) and the function $\mathrm{b}_{\mathrm{1hop}}$ in (\ref{eq:obj5.2}). It can be observed from the analysis in the previous sections that the new objective function must be a multi-concave function of the system variables as long as the function $\mathrm{g}$ described in (\ref{eq:g_fun}) is a multi-convex function. This however only requires that each estimated data symbol $\hat{d}^{\left(m\right)}$ is a multi-affine function of the system variables. Therefore, we may conclude that for a system in which the estimated data symbols are multi-affine functions of the system variables, the sum rate maximization problem can be equivalently formulated as a multi-convex optimization problem.

In this paper, we only considered the case where each MS receives a single desired data symbol from the corresponding BS. If more than one data symbol is desired by each MS, one can expect that the estimated data symbols at a MS are superposed by colored noise. Therefore, a pre-whitening filter is required at each MS to decorrelate the noise signals, which results in that the estimated data symbols are no longer multi-affine functions of the system variables. However, if this correlation among the noise signals is ignored and the received data symbols are decoded symbol-wise, an approximate sum rate can still be maximized using the proposed algorithm.

\section{Numerical results}
\label{sec:7}
In the following simulation results, the performance of the proposed sum rate maximization algorithm is evaluated in a cellular scenario with $K=2$ cells, $M=3$ MSs per cell, $N_{\mathrm{B}}=3$ antennas at each BS, $R=4$ relays, and $N_{\mathrm{R}}=N_{\mathrm{M}}=2$ antennas at each relay and MS. The two-hop transmission scheme is applied. Concerning the channel model, we employ an i.i.d. complex Gaussian channel model with the average channel gain being normalized to one. To assess the performance of our proposed algorithm, two reference schemes are considered. Firstly, an IA algorithm is considered where the tuple $\mathbb{V}$ of the transmit filters, the tuple $\mathbb{U}$ of the receive filters, and the tuple $\mathbb{G}$ of the relay processing matrices are alternatingly optimized to minimize the total interference leakage in the system. The considered IA algorithm is a direct extension of the interference leakage minimization algorithm proposed in \cite{GCJ11} to a multiuser relay scenario. The second 
reference scheme is the sum MSE minimization algorithm which minimizes the sum 
MSE by alternatingly optimizing $\mathbb{V}$, $\mathbb{G}$, and $\mathbb{U}$ \cite{SSBHU09,SLTW10,MXFWNP10,GAWK13}.

\begin{figure}[!t]
  \centering
  \psfrag{002}[c][c]{sum rate in bits$/$channel use}
  \psfrag{001}[ct][ct]{$10\textrm{log}_{10}\left(\gamma_\mathrm{PSNR}\right)$ in dB}
  \psfrag{003}[rb][rb]{sum rate max.}
  \psfrag{004}[rb][rb]{sum MSE min.}
  \psfrag{005}[rb][rb]{IA}
  \includegraphics[width=3.5in]{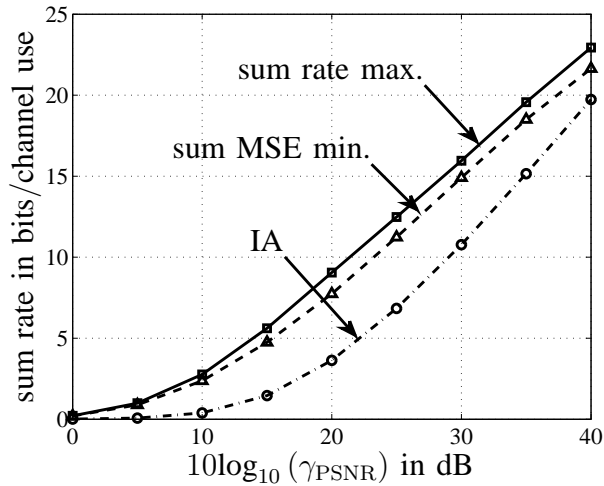}\\
  \caption{Average sum rate per time slot as a function of the pseudo SNR $\gamma_{\mathrm{pSNR}}$ in decibel for a scenario with $K=2$, $M=N_{\mathrm{B}}=3$, $R=4$, and $N_{\mathrm{R}}=N_{\mathrm{M}}=2$}\label{fig:result}
\end{figure}
Firstly, the achieved sum rate per time slot is considered as a performance measure. The performance of the proposed algorithm is considered as a function of the pseudo SNR which is defined as the ratio of the sum transmit power of all the BSs and relays to the noise variance $\sigma^{2}$, i.e.,
\begin{equation}\label{eq:PSNR}
    \gamma_{\mathrm{PSNR}}=\frac{P_{\mathrm{B}}+P_{\mathrm{R}}}{\sigma^{2}}.
\end{equation}
Figure \ref{fig:result} shows the performances of the three considered algorithms averaged over many different channel snapshots. It can be seen from Fig. \ref{fig:result} that the IA algorithm performs poorly as compared to the other two algorithms at low to moderate pseudo SNRs. On the one hand, the IA algorithm does not consider noise reduction. On the other hand, the IA algorithm does not intend to improve the received powers of the useful signals when minimizing the interferences. That is to say, the IA algorithm does not maximize the received SNRs at the MSs. In the pseudo SNR region shown in Fig. \ref{fig:result}, both the sum MSE minimization and the sum rate maximization algorithms achieve superior performance as compared to the IA algorithm. However, the sum rate maximization algorithm outperforms the sum MSE minimization algorithm on average. This shows that minimizing the sum MSE does not necessarily achieve high sum rates. At high pseudo SNRs, interferences become more harmful. As IA aims at 
perfectly nullifying all the interferences, the sum rates achieved by the IA algorithm increase approximately linearly with the pseudo SNRs and the slope is related to the achieved degrees of freedom (DoFs). Furthermore, if the sum MSE minimization and the sum rate maximization algorithms are able to find the global optima, all three curves should have the same slope at high pseudo SNRs. However, as the total available power increases, the feasible region described by the constraint sets of (\ref{eq:con1.1}) and (\ref{eq:con1.2}) enlarges and this complicates the search for a good local optimum for both the sum MSE minimization and the sum rate maximization algorithms. As a result, both algorithms cannot achieve the same DoFs as the IA algorithm.
\begin{figure}[!t]
  \centering
  \psfrag{001}[ct][ct]{sum rate in bits$/$channel use}
  \psfrag{002}[c][c]{probability density}
  \psfrag{003}[r][r]{sum rate max.}
  \psfrag{004}[rb][rb]{sum MSE min.}
  \psfrag{005}[b][b]{IA}
  \includegraphics[width=3.5in]{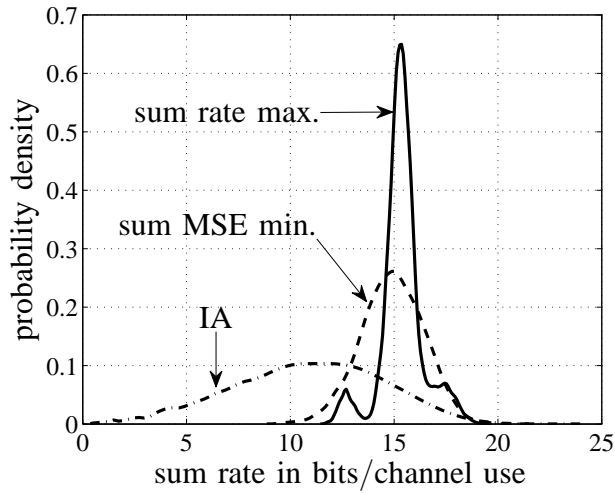}\\
  \caption{Probability density of the average sum rate per time slot at the pseudo SNR $\gamma_{\mathrm{pSNR}}=30\,\textrm{dB}$ for a scenario with $K=2$, $M=N_{\mathrm{B}}=3$, $R=4$, and $N_{\mathrm{R}}=N_{\mathrm{M}}=2$}\label{fig:distribution}
\end{figure}

Next we will take a closer look at the convergence of the proposed sum rate maximization algorithm. In Fig. \ref{fig:distribution}, the approximated probability density of the sum rates achieved by the proposed sum rate maximization algorithm, the sum MSE minimization algorithm, and the IA algorithm at a pseudo SNR of $30\,\textrm{dB}$ are shown. One can observe that the IA algorithm sometimes achieves a high sum rate but the average performance remains low. This implies that the SNR at each MS may vary across a wide range depending on the channel realization. The performance of the sum MSE minimization algorithm is more stable than that of the IA algorithm because the received useful signal powers are forced close to the transmit signal powers, i.e., the gains of the useful links are close to one. Finally, the proposed sum rate maximization algorithm achieves the highest average sum rate with the smallest variance among the three considered algorithms in this case. For a randomly given channel realization, 
the algorithm converges, with high probability, to a solution which achieves a sum rate in the range between $14$ bits per channel use and $17$ bits per channel use. However, for some channel realizations, the algorithm may also converge to solutions achieving a sum rate of about $13$ bits per channel use or $18$ bits per channel use. The reason is that the sum rate maximization algorithm is not guaranteed to achieve a global maximum. If the auxiliary variables $\mathbf{w}$ and $\mathbf{t}$ are fixed, the optimization problem of (\ref{eq:obj6})--(\ref{eq:con6.2}) is similar to a weighted sum MSE minimization problem. How to find the global optimum for such a problem is still an open question. In fact, alternatingly adapting the sets of optimization variables may result in that one or several users are turned off. In our simulation results for instance, it may happen that zero, one, or even two of the six MSs are turned off depending on the pseudo SNR and the channel realizations. Because of this, the IA 
algorithm 
can even outperform the proposed sum rate maximization algorithm at very high pseudo SNRs.

\begin{figure}[!t]
  \centering
  \psfrag{002}[c][c]{sum rate in bits$/$channel use}
  \psfrag{001}[ct][ct]{number of iterations}
  \psfrag{003}[cb][cb]{sum rate max.}
  \psfrag{004}[cb][cb]{sum MSE min.}
  \psfrag{005}[cb][cb]{IA}
  \includegraphics[width=3.5in]{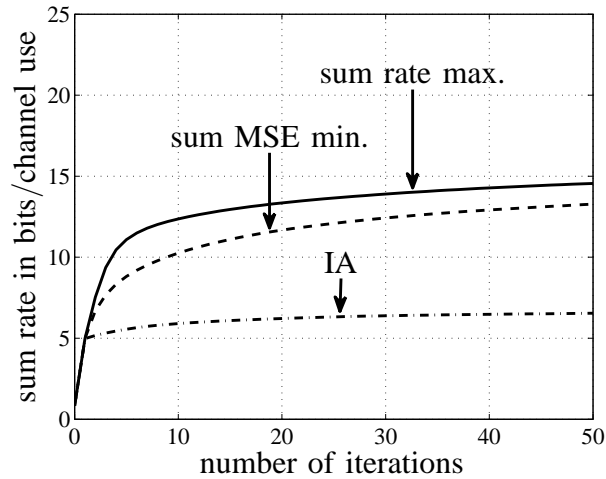}\\
  \caption{Average sum rate versus the number of iterations at the pseudo SNR $\gamma_{\mathrm{pSNR}}=30\,\textrm{dB}$ for a scenario with $K=2$, $M=N_{\mathrm{B}}=3$, $R=4$, and $N_{\mathrm{R}}=N_{\mathrm{M}}=2$}\label{fig:convergence}
\end{figure}
Figure \ref{fig:convergence} shows the average sum rate versus the number of iterations of the considered algorithms for the first 50 iterations at a pseudo SNR of $30\,\textrm{dB}$. Since reducing interferences does not guarantee an increment of the sum rate, the sum rate achieved by the IA algorithm converges. The average sum rate achieved by the sum MSE minimization algorithm slowly increases with the number of iterations. As compared to the sum MSE algorithm, the proposed sum rate maximization algorithm not only converges faster but also converges to a higher sum rate on average. The main reason is that the auxiliary variables $\mathbf{w}$ and $\mathbf{t}$ are adapted in every iteration to help maximizing the sum rate. Furthermore, since these auxiliary variables can be calculated in 
closed form, our sum rate maximization algorithm is comparable with the sum MSE minimization algorithms in terms of its computational complexity, which is mainly determined by the quadratic optimization tools used for optimizing the transmit filters at the BSs and the relay processing matrices.
\section{Conclusion}
\label{sec:8}
In this paper, the sum rate maximization problem in cellular networks is considered. It is shown that by adding two sets of auxiliary variables, this problem can be formulated as a multi-convex optimization problem. The property of multi-convexity in the new formulation makes it possible to find a local optimum using a low complexity iterative algorithm. The new proposed multi-convex formulation is not limited to our considered scenario, but it can be applied to many multiuser wireless system in which the estimated data symbols are multi-affine functions of the system variables.


%


\section*{Acknowledgment}
This work is supported by Deutsche Forschungsgemeinschaft (DFG), grants No. WE2825/11-1 and KL907/5-1. Hussein Al-Shatri and Anja Klein performed this work in the context of the DFG funded Collaborative Research Center (SFB) 1053 Multi-Mechanism-Adaptation for the Future Internet (MAKI).
\bibliographystyle{IEEEtran}

\end{document}